\documentclass[aps,onecolumn]{revtex4}
\usepackage{graphicx}
\usepackage{epsfig}
\usepackage{epstopdf}
\usepackage{amsfonts}
\usepackage{amssymb}
\usepackage{amsbsy}
\usepackage{amsmath}
\usepackage{mathrsfs}
\usepackage{latexsym}
\usepackage{natbib}
\usepackage{bm}
\usepackage{color}
\usepackage[a4paper, total={6.5in, 9.8in}]{geometry}

\DeclareMathOperator{\atanh}{atanh}

\usepackage{braket}
\usepackage{slashed}
\usepackage{pgfplots}
\usepackage{natbib}
\numberwithin{equation}{section}
\newtheorem{theorem}{Theorem}

\usepackage{tikz}
\usetikzlibrary{shapes,arrows,shadows}

\begin{document}

\title{Solitons in curved spacetime}

\author{Susobhan Mandal}
\email{sm17rs045@iiserkol.ac.in}

\affiliation{ Department of Physical Sciences,\\ 
Indian Institute of Science Education and Research Kolkata,\\
Mohanpur - 741 246, WB, India }

\date{\today}

\begin{abstract}
\begin{center}
\underline{\textbf{Abstract}}
\end{center}
Derrick's theorem is an important result that decides the existence of soliton configurations in field theories in different dimensions. It is proved using the extremization of finite energy of configurations under the scaling transformation. According to this theorem, the $2+1$ dimension is the critical dimension for the existence of solitons in scalar field theories without the gauge fields. In the 
present article, Derrick's theorem is extended in a generic curved spacetime in a covariant manner. Moreover, the existence of solitons in conformally flat spacetimes and spherically symmetric spacetimes is also shown using the approach presented in this article. Further, the approach shown in the present article in order to derive the soliton configurations is not restricted to a particular form of the field potential or curved spacetime.
\end{abstract}

\maketitle

\section{Introduction}
Soliton configurations play important roles in many physical phenomena, ranging from nuclear physics \cite{kalafatis1992soliton, goldflam1982soliton, ohno1986soliton, andrianov1988scalar, frank1991chiral, alkofer1996baryons} to gravitational physics \cite{rybakov1997solitons, mielke2002nontopological, ponglertsakul2016stability, kunz2013gravitating, franzin2018sine}. Solitons are the non-perturbative solution of field equations that lead to important physical effects \cite{torgrimsson2017dynamically, faddeev1978quantum, callan1991supersymmetric, kivshar1994gordon, kivshar1989dynamics, gordon1986theory, kaup1978solitons} in different quantum field theories, optical systems, condensed matter systems, and other physical systems. Unlike the Minkowski spacetime, the existence of soliton configurations in a generic curved spacetime is not shown yet systematically in the literature \cite{gonzalez2001scalar, franzin2018sine, radu2012spinning}. As a result, in order to show the existence of solitons in general relativity, a numerical approach is often used in the literature \cite{gonzalez2001scalar, franzin2018sine, radu2012spinning}. Derrick's theorem \cite{derrick1964comments} answers the question of the existence of soliton configurations in different dimensions in Minkowski spacetime, however, this theorem is not extended to a generic curved spacetime. The foremost theme of the present article is to extend Derrick's theorem in an arbitrary curved spacetime in a covariant manner. In this extended version of Derrick's theorem, the role of explicit dependence of the field potential on the spacetime coordinates in the existence of solitons is shown explicitly. Moreover, geometric properties like metric connection, curvature tensors also play important roles in deciding the possible form of field potentials in which soliton configurations can exist. Using this extension of Derrick's theorem, it is shown that the existence of solitons is shown explicitly in conformally flat spacetimes and spherically symmetric curved spacetimes. This is also applicable to other curved spacetimes. The above extension of Derrick's theorem in curved spacetime can show the existence of soliton configurations in gravitating systems \cite{semelin2001self, kunz2013gravitating} that give rise to new equilibrium configurations without gravitational collapse \cite{bednarek1998soliton, brito2001network, lee1987soliton} known as the soliton stars. Moreover, the extension of Derrick's theorem in curved spacetime is also important in order to know about the formation of solitons by the degenerate plasma inside the neutron stars, and white-dwarfs \cite{hossain2019revisiting, el2020oblique, berezhiani2015electromagnetic}.
 
\section{Derrick's theorem}
In this section, we present Derrick's theorem as a preliminary material for our later studies. In the present article, we used the general relativity (GR) convention of metric signature $(-,+,\ldots,+)$.
\begin{theorem}
Let $\{\phi^{A}\}$ be a set of scalar fields in $D+1$-dimensional Minkowski spacetime. If the action of this theory is given by the following
\begin{equation}
S=-\int d^{D+1}x\Big[\frac{1}{2}\partial_{\mu}\phi^{A}(x)\partial^{\mu}\phi^{A}(x)+U[\{\phi^{A}\}]\Big],
\end{equation}
such that $U[\{\phi^{A}\}]\geq0$ and equals to zero only for vacuum states, then i) there may exist solitons for $D=1$, ii) there may exist solitons for $D=2$ only if $U[\{\phi^{A}\}]=0$ identically and iii) there do not exist any soliton for $D>2$.
\end{theorem}

For a set of static configurations, the energy of the system is given by
\begin{equation}\label{enrgy-expression}
\mathcal{E}=\int d^{D}x\Big[\frac{1}{2}\partial_{i}\phi^{A}(x)\partial_{i}\phi^{A}(x)+U[\{\phi^{A}\}]\Big],
\end{equation}
which is positive definite. If we consider the set of static soliton solutions of the Euler-Lagrange equations, then the energy given by the above expression would be finite and extremum. This also implies that under any continuous transformation, the energy corresponding to the static soliton solutions of the Euler-Lagrange equation must be extremum. In order to check that, we choose the scale transformation given by $\phi^{A}(x)\rightarrow\phi_{\lambda}^{A}(x)=\phi^{A}(\lambda x)$. Under this transformation, the expression of energy behaves as
\begin{equation}
\begin{split}
\mathcal{E}_{\lambda} & =\int d^{D}x\Big[\frac{1}{2}\frac{\partial}{\partial x^{i}}\phi^{A}(\lambda x)\frac{\partial}{\partial x^{i}}\phi^{A}(\lambda x)+U[\{\phi^{A}(\lambda x)\}]\Big]\\
 & =\Big[\lambda^{2-D} \int d^{D}y \ \frac{1}{2}\frac{\partial}{\partial y^{i}}\phi^{A}(y)\frac{\partial}{\partial y^{i}}\phi^{A}(y)+\lambda^{-d} \int d^{D}y \ U[\{\phi^{A}(y)\}]\Big]\\
 & \equiv\lambda^{2-D}T+\lambda^{-D}V.
\end{split}
\end{equation}
According to the extremum condition, we expect $\frac{d}{d\lambda}\mathcal{E}_{\lambda}\Big|_{\lambda=1}=0$. This leads to the following constraint
\begin{equation}
(2-D)T-DV=0,
\end{equation}
where $T$ and $V$ are the kinetic and potential parts of the expression of energy in (\ref{enrgy-expression}), respectively. The above constraint indeed shows that solitons exist i) for $D=1$ if $T=V$, ii) for $D=2$ if $V=0$ and iii) solitons do not exist for $D=3$. However, the above conclusion does not hold if the potential in the action $U[x,\{\phi_{\lambda}^{A}(\lambda x)\}]$ depends on the spatial coordinates explicitly. As a consequence, Derrick's theorem \cite{derrick1964comments} is not valid in such situations \cite{bazeia2003new}. Further, Derrick's theorem is not valid in curved spacetime due to the coupling between matter and the metric of the background geometry. In \cite{palmer1979derrick, carloni2019derrick, alestas2019evading}, Derrick's theorem is extended to field theories in a class of curved spacetimes. The existence of soliton configuration is also shown numerically in \cite{gonzalez2001scalar}. However, these approaches are not general enough to find the existence of solitons in a generic curved spacetime.  

\section{Extending Derrick's theorem to curved spacetime}
In order to show the existence of solitons in curved spacetime, we consider the following minimally coupled scalar field theory in $D+1$ dimension with metric $g_{\mu\nu}(x)$ \textit{w.r.t} Cartesian coordinates
\begin{equation}\label{curved action}
\begin{split}
S & =-\int\sqrt{-g(x)}d^{D+1}x\Big[\frac{1}{2}g^{\mu\nu}(x)\partial_{\mu}\phi^{A}(x)\partial_{\nu}\phi^{A}(x)+U[\{\phi^{A}(x)\},x]\Big],
\end{split}
\end{equation}
where the metric signature is considered according to GR convention. The field potential in the above action also depends on the spacetime coordinates explicitly apart from the field variables. For a static configuration, the energy is given by the following expression
\begin{equation}\label{energy curved}
\begin{split}
\mathcal{E}(t) & =\int_{\Sigma_{t}}\sqrt{-g(x)}d^{D}x\Big[\frac{1}{2}g^{ij}(x)\partial_{i}\phi^{A}(x)\partial_{j}\phi^{A}(x)+U[\{\phi^{A}(x)\},x]\Big],
\end{split}
\end{equation}
where $g^{ij}(x)$ are positive definite functions. In order to emphasize on the static configuration, we consider the time-like coordinate to be constant in the metric components and the above integration is done on a spacelike hypersurface $\Sigma_{t}$.
  
In this section, we derive the conditions in a covariant manner under which a soliton configuration can exist in a generic curved spacetime. Under the scale transformation of fields, the expression for the energy is given by
\begin{equation}
\begin{split}
\mathcal{E}_{\lambda}(t) & =\int_{\Sigma_{t}}\sqrt{-g(t,x)}d^{D}x\Big[\frac{1}{2}g^{ij}(t,x)\frac{\partial}{\partial x^{i}}\phi^{A}(t,\lambda x)\frac{\partial}{\partial x^{j}}\phi^{A}(t,\lambda x)+U[\{\phi^{A}(t,\lambda x)\},x]\Big]\\
=\lambda^{-D} & \int_{\Sigma_{t}}\sqrt{-g\left(t,\frac{y}{\lambda}\right)}d^{D}y\Big[\frac{\lambda^{2}}{2}g^{ij}\left(t,\frac{y}{\lambda}\right)\frac{\partial}{\partial y^{i}}\phi^{A}(t,y)\frac{\partial}{\partial y^{j}}\phi^{A}(t,y)+U\left(\{\phi^{A}(t,y)\},\frac{y}{\lambda}\right)\Big].
\end{split}
\end{equation}  
Here onwards we do not write coordinate-time dependences since we restrict our discussion to static solitons in the present article. Using the relations $\partial_{k}g^{ij}(y)=-2\Gamma_{ \ \ k}^{(i \ \ j)}(y)$, and $\partial_{k}\log\sqrt{-g(y)}=\Gamma_{ \ \mu k}^{\mu}(y)$, the extremization condition $\frac{d\mathcal{E}_{\lambda}(t)}{d\lambda}\Big|_{\lambda=1}=0$ can be expressed as
\begin{equation}\label{curved constraint 1}
\begin{split}
\int_{\Sigma_{t}} & d^{D}y\Bigg[\frac{1}{2}\mathcal{K}_{D}^{ij}(y)\frac{\partial\phi^{A}(y)}{\partial y^{i}}\frac{\partial\phi^{A}(y)}{\partial y^{j}}-(D+y^{k}\Gamma_{ \ \mu k}^{\mu}(y))U[\{\phi^{A}(y)\},y]\\
 & -y^{k}\frac{\partial U}{\partial y^{k}}[\{\phi^{A}(y)\},y]\Bigg]\sqrt{-g(y)}=0,
\end{split}
\end{equation}
where
\begin{equation}
\mathcal{K}_{D}^{ij}(y)=\Big[(2-D)-\Gamma_{ \ \mu k}^{\mu}(y)y^{k}\Big]g^{ij}(y)+2y^{k}\Gamma_{ \ \ k}^{(i \ \ j)}(y).
\end{equation}
The above relation is a non-trivial relation, as it depends on the geometric quantities like the metric and the Christoffel symbols $\Gamma_{ \ \mu\nu}^{\lambda}$ apart from the spatial dimension $D$. On the other hand, the minimization of energy requires $\frac{d^{2}\mathcal{E}_{\lambda}(t)}{d\lambda^{2}}\Big|_{\lambda=1}>0$, which demands the following constraint to be satisfied
\begin{equation}\label{curved constraint 2}
\begin{split}
\int_{\Sigma_{t}}\sqrt{-g(y)}d^{D}y\Big[ & \mathcal{P}_{D}^{ij}(y)\frac{\partial\phi^{A}(y)}{\partial y^{i}}\frac{\partial\phi^{A}(y)}{\partial y^{j}}+\mathcal{W}[\phi^{A}(y),y]\Big]>0, 
\end{split}
\end{equation}
where the expressions of $\mathcal{P}_{D}^{ij}(y)$, and $\mathcal{W}[\phi^{A}(y),y]$ are given by
\begin{equation}
\begin{split}
\mathcal{P}_{D}^{ij}(y) & =\frac{(D-1)(D-2)}{2}g^{ij}(y)+(D-1)y^{k}\Big[-2\Gamma_{ \ \ k}^{(i \ \ j)}(y)+\Gamma_{ \ \mu k}^{\mu}(t,y)g^{ij}(y)\Big]\\
 & +\frac{1}{2}y^{k}y^{l}\Big[-2\partial_{l}\Gamma_{ \ \ k}^{(i \ \ j)}(y)-4\Gamma_{ \ \mu k}^{\mu}(y)\Gamma_{ \ \ l}^{(i \ \ j)}(y)+\partial_{l}\Gamma_{ \ \mu k}^{\mu}(t,y)g^{ij}(y)\\
 & +\Gamma_{ \ \mu k}^{\mu}(y)\Gamma_{ \ \mu l}^{\mu}(t,y)g^{ij}(y)\Big],
\end{split}
\end{equation}
and
\begin{equation}
\begin{split}
\mathcal{W} & =D(D+1)U+2(D+1)y^{k}\frac{\partial U}{\partial y^{k}}+y^{k}y^{l}\frac{\partial^{2}U}{\partial y^{k}\partial y^{l}}+(D+2)y^{k}\Gamma_{ \ \mu k}^{\mu}U\\
 & +y^{k}y^{l}\Gamma_{ \ \mu k}^{\mu}\Gamma_{ \ \mu k}^{\mu}U+y^{k}\Gamma_{ \ \mu k}^{\mu}y^{l}\frac{\partial U}{\partial y^{l}}+y^{k}y^{l}\frac{\partial\Gamma_{ \ \mu k}^{\mu}}{\partial y^{l}}U.
\end{split}
\end{equation}
In a different coordinate system, $D$ in (\ref{curved constraint 1}) and (\ref{curved constraint 2}) must be replaced by $d$ where $d$ is the number of spatial coordinates which scales linearly under the scaling of Minkowski coordinates. For example, in a spherical symmetric geometry, $d=1$ since radial coordinate is the only such spatial coordinate. The conditions (\ref{curved constraint 1}) and (\ref{curved constraint 2}) can also be derived similarly for field theories in curved spacetime with non-minimal couplings between fields and curvature.

Like the extremization of the constraint (\ref{curved constraint 1}), the constraint (\ref{curved constraint 2}) due to the minimization of energy is also non-trivial since it depends also on the geometric quantities of the background spacetime manifold. Both the potential and kinetic parts in (\ref{curved constraint 2}) depend on the Christoffel symbols and their derivatives apart from the dimensionality. Hence, the constraints (\ref{curved constraint 1}) and (\ref{curved constraint 2}) show clearly that the existence of solitons in curved spacetime depends both on the geometry of the background spacetime manifold and dimensionality. Moreover, the existence of solitons in curved spacetime also depends on the form of field potential. The solutions of the constraint (\ref{curved constraint 1}) with the Euler-Lagrange equations and (\ref{curved constraint 2}) give the extension of Derrick's theorem in a generic curved spacetime in a covariant manner. The results in the equations (\ref{curved constraint 1}) and (\ref{curved constraint 2}) can also be derived similarly in a different coordinate system.

\section{Solitons in conformally flat spacetimes}
Conformally flat spacetimes are often used in different areas of GR \cite{romero2012conformally, gron2011frw1, gron2011frw2, mishra2020note}. Further, a significant number of physically interesting spacetimes predicted by GR belong to this class. Hence, it is important to know about the criteria under which solitons can exist in this class of spacetimes.

\subsection{Solitons in $1+1$-dimensional curved spacetime}
It can be checked easily that the metric of a given $1+1$-dimensional curved spacetime can always be expressed as
\begin{equation}
ds^{2}=\Omega(t,y)(-dt^{2}+dy^{2}),
\end{equation}
where $\Omega(t,y)$ is the conformal factor which may not always be positive definite for a time-like coordinate $t$ and spacelike coordinate $y$. Hence, we obtain the following relations
\begin{equation}
\begin{split}
\partial_{k}g^{ij}(t,y) & =-2\Gamma_{ \ \ k}^{(i \ \ j)}(t,y)=-\delta^{ij}\frac{1}{\Omega^{2}(t,y)}\partial_{y}\Omega(t,y)\\ 
\Gamma_{ \ \mu k}^{\mu} & =\frac{1}{\Omega(t,y)}\partial_{y}\Omega(t,y).
\end{split}
\end{equation}
Now onwards, we omit the time coordinate since we restrict our discussion to static configurations. Using the above two relations, the constraints (\ref{curved constraint 1}) and (\ref{curved constraint 2}) become
\begin{equation}\label{conformal constraint 1}
\int_{\Sigma_{t}}dy\Big[\frac{1}{2}\frac{\partial\phi^{A}}{\partial y}\frac{\partial\phi^{A}}{\partial y}-\Omega U-y\Omega\frac{\partial U}{\partial y}-yU\frac{d\Omega}{dy}\Big]=0,
\end{equation}
and 
\begin{equation}\label{conformal constraint 2}
\begin{split}
\int_{\Sigma_{t}} & \Omega dy \Big[2U+4y\frac{\partial U}{\partial y}+2y^{2}\frac{\partial^{2}U}{\partial y^{2}}+3y\frac{d\log\Omega}{dy}U\\
 & +y^{2}\left(\frac{d\log\Omega}{dy}\right)^{2}U+y^{2}\frac{d\log\Omega}{dy}\frac{\partial U}{\partial y}+y^{2}\frac{d^{2}\log\Omega}{dy^{2}}U\Big]>0,
\end{split}
\end{equation}
respectively in a generic $1+1$-dimensional curved spacetime. The above two conditions can be satisfied clearly in $1+1$-dimensional curved spacetime, hence, there exist solitons in $1+1$-dimensional curved spacetimes despite the presence of non-trivial conformal factor $\Omega(t,y)$. In a generic $1+1$-dimensional curved spacetime, the expression (\ref{energy curved}) reduces to the following expression
\begin{equation}
\begin{split}
\mathcal{E}(t) & =\int_{\Sigma_{t}}dy\Big[\frac{1}{2}\frac{\partial\phi^{A}(y)}{\partial y}\frac{\partial\phi^{A}(y)}{\partial y}+\Omega(y)U[\{\phi^{A}(y)\},y]\Big]\\
 & =\int_{\Sigma_{t}}dy\Big[2\Omega(y)U[\{\phi^{A}(y)\},y]+y\Omega(y)\frac{\partial U[\{\phi^{A}(y)\},y]}{\partial y}+yU[\{\phi^{A}(y)\},y]\frac{d\Omega(y)}{dy}\Big].
\end{split} 
\end{equation}
In order to be a finite-energy configuration, both $\frac{d\phi^{A}}{dy}$, and $\Omega U$ must vanish at the boundary.  

Now we solve the Euler-Lagrange equation for a single scalar field theory in a consistent manner such that the condition (\ref{conformal constraint 1}) holds. The Euler Lagrange equation for a static configuration is given by
\begin{equation}
\frac{d^{2}\phi}{dy^{2}}=\Omega\frac{\partial U}{\partial\phi}\implies\frac{d}{dy}\left(\frac{d\phi}{dy}\right)^{2}=2\Omega\frac{\partial U}{\partial\phi}\frac{d\phi}{dy},
\end{equation}
which follows from the action (\ref{curved action}) with a single scalar field. Now using the equation (\ref{conformal constraint 1}), the above equation can be expressed as
\begin{equation}\label{4.1}
\begin{split}
\frac{d}{dy}\Big[\Omega U & +y\Omega\frac{\partial U}{\partial y}+yU\frac{d\Omega}{dy}\Big]=\Omega\frac{\partial U}{\partial\phi}\frac{d\phi}{dy}\\
\implies U\frac{d\Omega}{dy} & +2\Omega\frac{\partial U}{\partial y}+y\frac{d}{dy}\left(\Omega\frac{\partial U}{\partial y}\right)+\frac{d}{dy}\left(yU\frac{d\Omega}{dy}\right)=0.
\end{split}
\end{equation}
For $\frac{\partial U}{\partial y}=0$, the above equation becomes
\begin{equation}
y\frac{dZ}{dy}=-2Z, \ Z=U\frac{d\Omega}{dy}.
\end{equation} 
This implies $U\frac{d\Omega}{dy}=\frac{\kappa}{y^{2}}$ where $\kappa$ is a constant and $\kappa$ is zero when $\frac{d\Omega}{dy}=0$. Therefore, the energy of this configuration is given by
\begin{equation}
\mathcal{E}=2\int_{-\infty}^{\infty}dy \ \Omega U =2\kappa\int_{-\infty}^{\infty}dy\frac{1}{y^{2}\frac{d\log\Omega}{dy}}.
\end{equation}
The above expression completely depends on the nature of the conformal factor, and it is only valid when $\frac{d\Omega}{dy}\neq0$. Hence, in order to have a finite-energy configuration, the above integral must be finite where $\Omega$ is a positive-definite function. For example, in Rindler spacetime, the above integral diverges as the conformal factor is given by $\Omega(y)=e^{2ay}$ where $a$ is the acceleration of the non-inertial observer. In order to find the suitable conformal factor which allows the existence of solitons, we must solve the following equation 
\begin{equation}
\frac{d\phi}{dy}=\pm\sqrt{2U(\phi)}\sqrt{\Omega+y\frac{d\Omega}{dy}},
\end{equation}
which follows from (\ref{conformal constraint 1}). Then plug it in the equation $U(\phi)\frac{d\Omega}{dy}=\frac{\kappa}{y^{2}}$. For an example, let us consider the field potential $U(\phi)=\frac{\lambda}{2}(\phi^{2}-\phi_{0}^{2})^{2}$. As a result, from the above equation, we obtain the following solution
\begin{equation}\label{4.1.1}
\phi(y)=\mp\phi_{0}\tanh\left(\sqrt{\lambda}\phi_{0}\Lambda(y)+\beta\right),
\end{equation}
where $\beta=\mp\atanh\left(\frac{\phi(y=0)}{\phi_{0}}\right)$ is a constant and
\begin{equation}\label{4.1.2}
\Lambda(y)=\int_{0}^{y}\sqrt{\Omega+y'\frac{d\Omega}{dy'}}dy'.
\end{equation}
Plugging the expression in the equation $U(\phi)\frac{d\Omega}{dy}=\frac{\kappa}{y^{2}}$, we obtain the following second order differential equation in $\Lambda$
\begin{equation}
\begin{split}
\left(\frac{d\Lambda}{dy}\right)^{2} & =\frac{d}{dy}\Big[-\frac{\kappa}{\mathcal{V}[\Lambda]}+y\left(\frac{d\Lambda}{dy}\right)^{2}\Big]\implies\frac{d^{2}\Lambda}{dy^{2}}=-\frac{\kappa}{y\mathcal{V}^{2}[\Lambda]},
\end{split}
\end{equation}
where $\mathcal{V}[\Lambda(y)]=U(\phi(y))$ is given by
\begin{equation}
\mathcal{V}[\Lambda]=\frac{\kappa}{y\Big[\left(\frac{d\Lambda}{dy}\right)^{2}-\Omega\Big]}, \ \frac{d(y\Omega)}{dy}=\left(\frac{d\Lambda}{dy}\right)^{2}.
\end{equation}
Solving the above differential equation with the initial conditions $\Lambda(y=0)=0, \ \frac{d\Lambda(y=0)}{dy}=a$ where $a$ is a constant, we can find the conformal factor $\Omega(y)$ from the equation (\ref{4.1.2}) for which solitons can exist. We also need to provide $\Omega(y=0)$ in order to solve the above coupled differential equations.

Now we consider a different situation. Let us consider a field potential of the form $U[\phi(y),y]=f(y)V[\phi(y)]$, then the
equation (\ref{conformal constraint 1}) leads to the following equation
\begin{equation}\label{4.2}
\frac{d\phi}{dy}=\pm\sqrt{2V[\phi]}\sqrt{\Omega f+y\frac{d(f\Omega)}{dy}}.
\end{equation}
However, the Euler Lagrange equation gives the following relation
\begin{equation}\label{4.3}
\begin{split}
y\frac{d(f\Omega)}{dy} & \frac{d\log V}{d\phi}\frac{d\phi}{dy}+2\frac{d(f\Omega)}{dy}+2y\frac{df}{dy}\frac{d\Omega}{dy}+y\left(\Omega\frac{d^{2}f}{dy^{2}}+f\frac{d^{2}\Omega}{dy^{2}}\right)=0.
\end{split}
\end{equation}
Combining the equations (\ref{4.2}) and (\ref{4.3}), we obtain the following relation
\begin{equation}
-\sqrt{\frac{2}{V}}\frac{dV}{d\phi}=\pm\frac{2\frac{d(f\Omega)}{dy}+2y\frac{df}{dy}\frac{d\Omega}{dy}+y\left(\Omega\frac{d^{2}f}{dy^{2}}+f\frac{d^{2}\Omega}{dy^{2}}\right)}{y\frac{d(f\Omega)}{dy}\sqrt{f\Omega+y\frac{d(f\Omega)}{dy}}}.
\end{equation}
The above equation gives the solution $\phi(y)$ for a given function $f(y)$ and the conformal factor $\Omega(y)$. However, this solution must be consistent with the solution coming from (\ref{4.2}). This fixes the function $f(y)$ for a given conformal factor $\Omega(y)$. Plugging the solution of (\ref{4.2}) in (\ref{4.3}), we obtain a well-defined second-order ordinary differential equation which must be solved to obtain $f(y)$ for a given $\Omega(y)$. This is shown explicitly in the next subsection through an example. This solution is a soliton configuration provided the inequality in (\ref{conformal constraint 2}) is satisfied. In a similar manner, for other forms of the field potential $U[\phi(y),y]$, the equation (\ref{conformal constraint 1}) and (\ref{4.1}) must be solved in order to get a soliton configuration.

\subsection{Solitons in $D+1$-dimensional conformally flat spacetime} 
The metric $g_{\mu\nu}(x)$ in a $D+1$-dimensional conformally flat spacetime is given by $g_{\mu\nu}(x)=\Omega^{2}(x)\eta_{\mu\nu}$ where $\eta_{\mu\nu}$ is the Minkowski metric. Here we restrict our discussion to spherically symmetric solitons or in other words, soliton configurations that only depend on the radial coordinate. Hence, we can write the metric $\eta_{\mu\nu}$ in spherical coordinates. Moreover, we need the metric element $g^{rr}$, the conformal factor $\Omega^{2}(r)$, and $\sqrt{-g(x)}\propto\Omega^{D+1}(r)r^{D-1}$ in order to find the soliton configurations. It is quite easy to check the following relations
\begin{equation}
\begin{split}
\partial_{r}g^{rr} & =-2\Gamma_{ \ \ r}^{(r \ \ r)}=-\frac{2}{\Omega^{3}}\frac{d\Omega}{dr}\\
\Gamma_{ \ \mu r}^{\mu} & =\frac{d}{dr}\log(\Omega^{D+1}r^{D-1})=(D+1)\frac{d\log\Omega}{dr}+\frac{D-1}{r}.
\end{split}
\end{equation}
Therefore, the constraint in (\ref{curved constraint 1}) leads to the following equality
\begin{equation}\label{4.4}
\frac{1}{2}g_{1}(r)\left(\frac{d\phi}{dr}\right)^{2}=r\frac{\partial U}{\partial r}+g_{2}(r)U,
\end{equation}
where
\begin{equation}
\begin{split}
g_{1}(r) & =\left(\frac{2-D}{\Omega^{2}}-\frac{r(D-1)}{\Omega^{3}}\frac{d\Omega}{dr}\right)\\
g_{2}(r) & =\left(D+(D+1)r\frac{d\log\Omega}{dr}\right).
\end{split}
\end{equation}
On the other hand, the Euler-Lagrange equation for a static configuration becomes
\begin{equation}\label{4.5}
\begin{split}
\frac{d^{2}\phi}{dr^{2}} & +(D-1)g_{3}(r)\frac{d\phi}{dr}=\Omega^{2}\frac{\partial U}{\partial\phi}\\
\implies\frac{d}{dr}\left(\frac{d\phi}{dr}\right)^{2} & +2(D-1)g_{3}(r)\left(\frac{d\phi}{dr}\right)^{2}=2\Omega^{2}\frac{\partial U}{\partial\phi}\frac{d\phi}{dr},
\end{split}
\end{equation}
where
\begin{equation}
g_{3}(r)=\left(\frac{1}{r}+\frac{d\log\Omega}{dr}\right).
\end{equation}
Combining the equations (\ref{4.4}) and (\ref{4.5}), we obtain the following relation 
\begin{equation}
\begin{split}
\frac{d}{dr}\Big[\frac{r\frac{\partial U}{\partial r}+g_{2}(r)U}{g_{1}(r)}\Big] & +2(D-1)\frac{g_{3}(r)}{g_{1}(r)}\left(r\frac{\partial U}{\partial r}+g_{2}(r)U\right)=\Omega^{2}\frac{\partial U}{\partial\phi}\frac{d\phi}{dr}.
\end{split}
\end{equation}
As earlier, considering the field potential of the form $U[\phi(r),r]=h(r)G(\phi(r))$, the above expression reduces to
\begin{equation}\label{4.6}
\frac{dG}{d\phi}\left(\Omega^{2}g_{1}^{2}-\frac{g_{1}r}{h}\frac{dh}{dr}-g_{1}g_{2}\right)\sqrt{\frac{2}{g_{1}G}}=\pm\frac{K_{1}-K_{2}+K_{3}}{\sqrt{r\frac{dh}{dr}+g_{2}h}},
\end{equation}
where
\begin{equation}
\begin{split}
K_{1} & =g_{1}\left(\frac{d\log h}{dr}+\frac{r}{h}\frac{d^{2}h}{dr^{2}}+\frac{dg_{2}}{dr}+g_{2}\frac{d\log h}{dr}\right)\\
K_{2} & =\left(r\frac{d\log h}{dr}+g_{2}\right)\frac{dg_{1}}{dr}\\
K_{3} & =2(D-1)g_{3}g_{1}\left(r\frac{d\log h}{dr}+g_{2}\right).
\end{split}
\end{equation}
The expression of energy for such a static configuration is given by
\begin{equation}
\begin{split}
\mathcal{E} & =\frac{2\pi^{\frac{D}{2}}}{\Gamma\left(\frac{D}{2}\right)}\int_{0}^{\infty}dr (r\Omega)^{D-1}\Big[\frac{1}{2}\left(\frac{d\phi}{dr}\right)^{2}+\Omega^{2}U\Big]\\
 & =\frac{2\pi^{\frac{D}{2}}}{\Gamma\left(\frac{D}{2}\right)}\int_{0}^{\infty}dr (r\Omega)^{D-1}\Big[\frac{r}{g_{1}}\frac{\partial U}{\partial r}+\frac{g_{2}}{g_{1}}U+\Omega^{2}U\Big].
\end{split} 
\end{equation}
The expression of energy must be finite for a soliton configuration and it also must satisfy the inequality in (\ref{curved constraint 2}). Moreover, a soliton configuration exists only if the solution from (\ref{4.6}) is equivalent to solution coming from the equation (\ref{4.4}) which can be expressed as
\begin{equation}\label{4.7}
\int\frac{d\phi}{\sqrt{2G(\phi)}}=\pm\int dr\sqrt{\frac{r}{g_{1}}\frac{dh}{dr}+\frac{g_{2}}{g_{1}}h}.
\end{equation}
It is important here to note that the equation (\ref{4.6}) becomes an identity for $D=1$ when both $h(r)$ and $\Omega(r)$ are chosen to be constant since the both sides of the equation vanish identically. Then the equation (\ref{4.7}) gives the solution of soliton configuration. For example, if we choose $G(\phi)=\frac{\lambda}{2}(\phi^{2}-\phi_{0}^{2})^{2}$, then the solution of the equation (\ref{4.7}) is given by
\begin{equation}\label{4.8}
\phi(r)=\mp\phi_{0}\tanh\left(\sqrt{\lambda}\phi_{0}\chi(r)+\alpha\right), 
\end{equation}
where $\alpha=\mp\atanh\left(\frac{\phi(r=0)}{\phi_{0}}\right)$ is a constant and
\begin{equation}\label{4.8.1}
\chi(r)=\int_{0}^{r}dr'\sqrt{\frac{r'}{g_{1}}\frac{dh}{dr'}+\frac{g_{2}}{g_{1}}h}.
\end{equation}
The above expression of $\chi$ is well-defined provided the term $\left(\frac{r}{g_{1}}\frac{dh}{dr}+\frac{g_{2}}{g_{1}}h\right)$ is positive-definite. Plugging the expression (\ref{4.8}) into the equation (\ref{4.6}), we obtain the following relation
\begin{equation}\label{4.9}
4\sqrt{\lambda}\phi(r)\Big[\Omega^{2}h-\left(\frac{d\chi}{dr}\right)^{2}\Big]=\pm h\frac{J_{1}-J_{2}+J_{3}}{g_{1}^{2}\left(\frac{d\chi}{dr}\right)^{2}},
\end{equation}
where $J_{1}, J_{2}, J_{3}$ are given by the following expressions
\begin{equation}
\begin{split}
J_{1} & =\frac{1}{g_{1}h}\frac{d}{dr}\left(g_{1}\left(\frac{d\chi}{dr}\right)^{2}\right), \ J_{2}=\frac{1}{g_{1}h}\left(\frac{d\chi}{dr}\right)^{2}\frac{dg_{1}}{dr}, \ J_{3}=2(D-1)\frac{g_{3}}{h}\left(\frac{d\chi}{dr}\right)^{2}.
\end{split}
\end{equation}
Defining $\mathcal{G}=\frac{1}{h}\left(\frac{d\chi}{dr}\right)^{2}$, we obtain the following system of coupled first-order ordinary differential equations in $\mathcal{G}$ 
\begin{equation}\label{4.9.1}
\begin{split}
4\sqrt{\lambda}\phi(r) & \Big[\Omega^{2}-\mathcal{G}^{2}\Big]=\pm\frac{J_{1}-J_{2}+J_{3}}{g_{1}^{2}\mathcal{G}^{2}}\\
\frac{dh}{dr} & =\frac{h}{r}(g_{1}\mathcal{G}-g_{2}), \ \left(\frac{d\chi}{dr}\right)^{2}=h\mathcal{G},
\end{split}
\end{equation}
where $J_{1},J_{2},J_{3}$ can now be expressed as
\begin{equation}
\begin{split}
J_{1} & =\left(\frac{d\mathcal{G}}{dr}+\mathcal{G}\frac{d\log g_{1}}{dr}\right)+\frac{\mathcal{G}}{r}(g_{1}\mathcal{G}-g_{2})\\
J_{2} & =\mathcal{G}\frac{d\log g_{1}}{dr}, \ J_{3}=2(D-1)\mathcal{G}g_{3}.
\end{split}
\end{equation}
As a result, given a conformal factor $\Omega(r)$, the above set of coupled ordinary differential equations can be solved in principle given a suitable initial condition on $\{\phi, \chi, \mathcal{G}, h\}$ such that the conditions (\ref{curved constraint 1}) and (\ref{curved constraint 2}) are satisfied. The solutions of this system of differential equations give the functional form of $h(r)$ and $\chi(r)$. As a result, we also obtain the soliton configuration since $\phi(r)$ depends directly on $\chi(r)$. One of the solutions of (\ref{4.9.1}) is $\Omega=1, \mathcal{G}=\Omega, \chi=\frac{c_{1}}{r}, h=\frac{c_{2}}{r^{4}}$ where $\frac{c_{1}^{2}}{c_{2}}=1$. For $c_{1}=1$, we reproduced the result in \cite{morris2021radially}. 

The approach presented here is general compared to the one mentioned in \cite{morris2021radially}. Moreover, our approach can be extended to other form of field potentials in curved spacetime.

\section{Solitons in a $3+1$-dimensional spherically symmetric spacetime}
In \cite{morris2021radially}, the existence of solitons in a spherically symmetric spacetime is shown for a class of field theories with some restricted solutions using the result in \cite{atmaja2014bogomol}. In this section, we derive the general condition using the results in (\ref{curved constraint 1}) and (\ref{curved constraint 2}) under which solitons can exist in a generic spherically symmetric spacetime. The line element of a generic spherically symmetric spacetime is given by
\begin{equation}
ds^{2}=-A(r)dt^{2}+B(r)dr^{2}+C(r)r^{2}(d\theta^{2}+\sin^{2}\theta d\varphi^{2}).
\end{equation} 
As earlier, we restrict our discussion to spherically symmetric soliton configuration. Hence, we need these radial coordinate-dependent functions $g^{rr}=\frac{1}{B}, \ \sqrt{-g}\propto r^{2}v(r), \ \Gamma_{ \ \mu r}^{\mu}=\frac{d\log v}{dr}+\frac{2}{r}, \ \partial_{r}g^{rr}=-2\Gamma_{ \ \ r}^{(r \ \ r)}=-\frac{1}{B^{2}}\frac{dB}{dr}$ where $v=\sqrt{AB}C$. Using the above-mentioned functions, the constraint in (\ref{curved constraint 1}) becomes
\begin{equation}\label{4.10}
\frac{S_{1}}{2}\left(\frac{d\phi}{dr}\right)^{2}-S_{2}U-r\frac{\partial U}{\partial r}=0,
\end{equation}
where $U[\phi(r),r]$ is the field potential and
\begin{equation}
S_{1}=-\frac{1}{B}\left(1+r\frac{d\log v}{dr}\right)+\frac{r}{B^{2}}\frac{dB}{dr}, \ S_{2}=\left(3+r\frac{d\log v}{dr}\right).
\end{equation}
On the other hand, the Euler-Lagrange equation for a static configuration is given by
\begin{equation}\label{4.11}
\begin{split}
\frac{d^{2}\phi}{dr^{2}} & +S_{3}\frac{d\phi}{dr}=B\frac{\partial U}{\partial\phi}\\
\implies\frac{d}{dr}\left(\frac{d\phi}{dr}\right)^{2} & +2S_{3}\left(\frac{d\phi}{dr}\right)^{2}=2B\frac{\partial U}{\partial\phi}\frac{d\phi}{dr},
\end{split}
\end{equation}
where $S_{3}=\frac{2}{r}+\frac{d\log v}{dr}-\frac{d\log B}{dr}$. The expression of energy for a static configuration is given by
\begin{equation}
\begin{split}
\mathcal{E} & =4\pi\int_{0}^{\infty}dr r^{2}v\Big[\frac{1}{2}\left(\frac{d\phi}{dr}\right)^{2}+U\Big]\\
 & =4\pi\int_{0}^{\infty}dr r^{2}v\Big[\frac{r}{S_{1}}\frac{\partial U}{\partial r}+\frac{S_{2}}{S_{1}}U+U\Big].
\end{split}
\end{equation}
Here we choose the field potential of a different form given by $U[\phi(r),r]=f_{1}(r)U_{1}[\phi]+f_{2}(r)U_{2}[\phi]$. As a result, the equation (\ref{4.10}) becomes
\begin{equation}\label{4.12}
\begin{split}
\frac{1}{2}\left(\frac{d\phi}{dr}\right)^{2} & =U_{1}[\phi]L_{1}(r)+U_{2}[\phi]L_{2}(r)\\
\implies\frac{1}{2}\left(\frac{d\bar{\phi}}{dr}\right)^{2} & =L_{1}(r)+\bar{U}[\bar{\phi}]L_{2}(r),
\end{split}
\end{equation}
where $(d\phi/d\bar{\phi})=\sqrt{U_{1}[\phi]}, \ \bar{U}[\bar{\phi}]=\frac{U_{2}[\phi]}{U_{1}[\phi]}$, and
\begin{equation}
L_{i}=\left(\frac{S_{2}f_{i}}{S_{1}}+\frac{r}{S_{1}}\frac{df_{i}}{dr}\right), \ i=1,2.
\end{equation}
The equation (\ref{4.12}) can be solved in general for $\bar{\phi}(r)$ which eventually gives the static field configurations $\phi(r)$. On the other hand, plugging the first equation of (\ref{4.12}) in (\ref{4.11}), we obtain the following relation
\begin{equation}\label{4.13}
\pm\mathcal{K}\sqrt{2(L_{1}+L_{2}\bar{U})}=\Big[\bar{U}_{1}\frac{dL_{1}}{dr}+\bar{U}_{2}\frac{dL_{2}}{dr}\Big]+2S_{3}(\bar{U}_{1}L_{1}+\bar{U}_{2}L_{2}),
\end{equation}
where $\bar{U}_{i}[\bar{\phi}]=U_{i}[\phi]$ for $i=1,2$ and
\begin{equation}
\mathcal{K}=B\left(f_{1}\frac{d\bar{U}_{1}}{d\bar{\phi}}+f_{2}\frac{d\bar{U}_{2}}{d\bar{\phi}}\right)-\left(L_{1}\frac{d\bar{U}_{1}}{d\bar{\phi}}+L_{2}\frac{d\bar{U}_{2}}{d\bar{\phi}}\right).
\end{equation}
The equation (\ref{4.13}) is a second-order differential equation in $f_{1}, f_{2}$ which follows from the following expression
\begin{equation}
\begin{split}
\bar{U}_{i}\left(\frac{dL_{i}}{dr}+2S_{3}L_{i}\right) & =\bar{U}_{i}\Big[f_{i}\Big[\frac{d}{dr}\left(\frac{S_{2}}{S_{1}}\right)+\frac{2S_{3}S_{2}}{S_{1}}\Big]+\frac{r}{S_{1}}\frac{d^{2}f_{i}}{dr^{2}}\\
 & +\left(\frac{S_{2}}{S_{1}}+\frac{d}{dr}\left(\frac{r}{S_{1}}\right)+\frac{2S_{3}r}{S_{1}}\right)\frac{df_{i}}{dr}\Big].
\end{split}  
\end{equation}
The $\pm$ signs in (\ref{4.13}) corresponds to kink and anti-kink solitons. As a result, we obtain the following system of first-order coupled differential equations
\begin{equation}\label{4.14}
\begin{split}
\frac{d\bar{\phi}}{dr}=\pm\sqrt{2(L_{1}+L_{2}\bar{U})}, & \ L_{i}=\frac{S_{2}f_{i}}{S_{1}}+\frac{r}{S_{1}}\frac{df_{i}}{dr},\\
\pm\mathcal{K}\sqrt{2(L_{1}+L_{2}\bar{U})} & =\Big[\bar{U}_{1}\frac{dL_{1}}{dr}+\bar{U}_{2}\frac{dL_{2}}{dr}\Big]+2S_{3}(\bar{U}_{1}L_{1}+\bar{U}_{2}L_{2}).
\end{split}
\end{equation}
This shows that given one of the functions $f_{1}$ or $f_{2}$, the above set of differential equations can be solved which results in the other function provided suitable initial conditions such that the conditions (\ref{curved constraint 1}) and (\ref{curved constraint 2}) are satisfied. This method can also be applicable if one of the functions $L_{1}$ or $L_{2}$ is provided. For $f_{2}=0$, one of the solution is given by the equations $\frac{d\log L_{1}}{dr}=-2S_{3}$ and $L_{1}=f_{1}B$ which implies $f_{1}\propto\frac{B}{r^{4}v^{2}}$ which reproduces two other results in \cite{morris2021radially}.

On the other hand, we also have the following relation
\begin{equation}
f_{2}\frac{dL_{1}}{dr}-f_{1}\frac{dL_{2}}{dr}=\mathcal{Z}\Big[\frac{S_{2}}{S_{1}}+\frac{d}{dr}\left(\frac{r}{S_{1}}\right)\Big]+\frac{r}{S_{1}}\frac{d\mathcal{Z}}{dr},
\end{equation}
where $\mathcal{Z}=f_{2}\frac{df_{1}}{dr}-f_{1}\frac{df_{2}}{dr}=\frac{S_{1}}{r}(f_{2}L_{1}-f_{1}L_{2})$. Therefore, given a radial function for $\mathcal{Z}$, the four coupled differential equations in (\ref{4.14}) can be solved exactly in order to obtain soliton configuration. In the similar manner, we can obtain soliton configurations for the field potentials of the form $U(\phi(r),r)=\sum_{i=1}^{n}f_{i}(r)U_{i}[\phi]$ in curved spacetime for $n\geq2$.

\section{Discussion}
In this article, we extended Derrick's theorem in a generic curved spacetime in a covariant manner. Moreover, we also discuss the existence of solitons in conformally flat spacetimes and spherically symmetric spacetimes, which are important for many reasons. Although our discussion is restricted mainly to these two classes of spacetimes in the present article, the method introduced here is general enough to find soliton configurations in other curved spacetimes. Moreover, unlike in \cite{morris2021radially}, here the constraints (\ref{curved constraint 1}, \ref{curved constraint 2}) on the soliton configurations are derived consistently following the scaling transformation of fields. These constraints depend on geometric quantities like metric connections and their derivatives. One may also note that our approach is not restricted to any particular class of field potentials. The existence of AdS solitons and their properties have been discussed in the literature \cite{anabalon2016hairy, khoury2012worldvolume, ovrut2012heterotic}. The criteria derived here for the existence of solitons in curved spacetime are useful for studying solitons in different aspects of GR. This is important in particular as the soliton configurations give rise to stable equilibrium configurations of boson stars \cite{friedberg1987scalar, kleihaus2012stable}.

\section{Acknowledgement}
SM wants to thank IISER Kolkata for supporting this work through a doctoral fellowship. 

\bibliographystyle{unsrt}
\bibliography{draft}

\end{document}